\documentclass[twocolumn,apj]{emulateapj}

\begin{document}

\title{Correlation of Hard X-Ray and White Light Emission in Solar Flares}

\author{Matej Kuhar\altaffilmark{1,2}, S\"am Krucker\altaffilmark{1,3}, Juan Carlos Mart\'{i}nez Oliveros\altaffilmark{3}, Marina Battaglia\altaffilmark{1}, Lucia Kleint\altaffilmark{1}, Diego Casadei\altaffilmark{1}, Hugh S. Hudson\altaffilmark{3,4}}
\altaffiltext{1}{University of Applied Sciences and Arts Northwestern Switzerland, Bahnhofstrasse 6, 5210 Windisch, Switzerland}
\altaffiltext{2}{Institute for Particle Physics, ETH Z\"{u}rich, 8093 Z\"{u}rich, Switzerland}
\altaffiltext{3}{Space Sciences Laboratory, University of California, Berkeley, CA 94720-7450, USA}
\altaffiltext{4}{School of Physics and Astronomy, University of Glasgow, Glasgow G12 8QQ, UK}

\begin{abstract}
A statistical study of the correlation between hard X-ray and white light emission in solar flares is performed in order to search for a link between flare-accelerated electrons and white light formation. We analyze 43 flares spanning GOES classes M and X using  observations from RHESSI (Reuven Ramaty High Energy Solar Spectroscopic Imager) and HMI (Helioseismic and Magnetic Imager). We calculate X-ray fluxes at 30 keV and  white light fluxes at 6173 $\textrm{\AA}$ summed over the hard X-ray flare ribbons with an integration time of 45 seconds around the peak hard-X ray time. We find a good correlation between hard X-ray fluxes and excess white light fluxes, with a highest correlation coefficient of 0.68 for photons with energy of 30 keV. Assuming the thick target model, a similar correlation is found between the deposited power by flare-accelerated electrons and the white light fluxes. The correlation coefficient is found to be largest for energy deposition by electrons above $\sim$50 keV. At higher electron energies the correlation decreases gradually while a rapid decrease is seen if the energy provided by low-energy electrons is added. This suggests that  flare-accelerated electrons of energy $\sim$ 50 keV are the main source for white light production. 

 \end{abstract}
 

\section{Introduction}
Solar flares are the most energetic phenomena on the Sun \citep[e.g., ][]{Benz08}. In a solar flare, large numbers of electrons, accelerated in the solar corona, precipitate to the lower layers of the solar atmosphere, where they deposit their energy in interactions with ambient gas. The energy is emitted over the whole electromagnetic spectrum, from radio waves to $\gamma$ rays \citep[e.g., ][]{Fletcher11}. Particularly interesting is the radiation in the white light (WL) range (visible continuum) because it contains a significant fraction of the total flare energy. The difficulty in detecting white light flares (WLFs) is that the enhancement in WL emission is relatively faint compared to the pre-flare level, typically ranging from a few percent up to several tens of percent in extreme cases. For more ordinary flares (below M5 GOES class), such a faint enhancement can be lost in the temporal and spatial intensity fluctuations of the photosphere. For this reason, it was long believed that white light flares (WLFs) are rare and exotic phenomena; however, recent observations suggest that WL emission can be detected in flares as weak as GOES C1.6 \citep[e.g., ][]{Hudson06}. \par

Many studies \citep[e.g.,][]{Hudson72, Rust75, Chen05, Chen06, Yan14} have reported a close correlation in space and time between HXR and WL emissions, which is a strong indication of the connection between non-thermal electron beams and WL formation.
There are two main mechanisms proposed for explaining the WL emission in solar flares: direct heating and radiative back-warming. In \textit{direct heating}, the same electrons that produce HXR emission via bremsstrahlung locally heat and ionize the medium. The WL continuum emission  is probably  produced by the recombination of hydrogen. Recent studies indeed show WL and HXR sources originating from the same volume \citep[]{MartinezOliveros12, Battaglia12, Krucker15}. The problem of the direct heating model is that there are very few electrons (because of the steepness of the non-thermal electron spectrum) with sufficient energy to penetrate to the lower chromosphere and photosphere, at least in the standard thick-target model of HXR emission. In  \textit{radiative back-warming}, electrons are stopped in the upper chromosphere, where the UV continuum radiation is emitted, which heats the deeper layers and results in enhanced continuum emission \citep[]{Aboudarham86, Machado89, Metcalf90}. \par
Generally, most of the previous studies on WLFs concentrated on individual events. In this paper, we analyze 43 WLFs spanning GOES classes M and X using observations with RHESSI (Reuven Ramaty High Energy Solar Spectroscopic Imager)  \citep{Lin02} and HMI (Helioseismic and Magnetic Imager) \citep{Scherrer12} in order to explore the connection between the flare-accelerated electrons and WL formation. The authors are aware of only one statistical study which analyzed the HXR and WL emission correlation with a comparable number of events \citep{Matthews03}. Their main results relevant for this study are: 1)  All flares above M8 class are WLFs, 2) There is no dependence of WL contrast on spectral index, and 3) There is a weak correlation between the contrast in WL and the  energies $>2\cdot10^{28}$ergs$^{-1}$, although with a large scatter. For lower energies, there is no obvious  trend. \par
In this article, we analyze the correlation between the HXR and WL emission in WLFs. In section \ref{section2} we provide  information about the instruments and data analysis steps of this study. Temporal, spatial and intensity relationships of HXR and WL fluxes are presented in section \ref{section3}. Additionally, the correlation between the power deposited by non-thermal electrons and WL fluxes is analyzed. In section \ref{section4}, we discuss the results and their implications.\\

\section{Observations}\label{section2}

Two instruments are used for the purposes of this study: HMI  and RHESSI.\par
HMI is one of the three instruments onboard the Solar Dynamic Observatory (SDO) \citep{Pesnell12}. It observes the Solar disk in the Fe I absorption line at 617.3 nm with a spatial resolution of $1.0''$ \citep{Scherrer12}. We use the standard level 1.5 continuum filtergram data, with no limb darkening correction applied and a time cadence of 45 seconds.     \par
RHESSI is an instrument designed for hard X-ray/gamma ray imaging and spectroscopy of solar flares. The best spatial resolution obtainable is $\sim 2.3''$, and the spectral resolution is $1-10$ keV FWHM in the operating energy range from 3 keV to 17 MeV \citep{Lin02}. With RHESSI, we are able to  match the integration time of HMI. \par
Here we report joint observations of RHESSI and HMI of 43 WLFs spanning GOES classes M and X and occurring between 2011 and 2014. All flares larger than GOES M5 class in the RHESSI database are included in this study.  Events below M5 class in our sample have at least one of the following properties: occurrence in 2011, which was the year we studied in most detail, occurrence near the limb, or an intense, short duration peak in HXR (flares with this property often show WL emission). For each flare, we computed the HXR flux at 30 keV, the HXR spectral index, the limb darkening correction factor for WL emission, the WL flux, the WL relative enhancement and the deposited energy by non-thermal electrons above 50 keV. These data, together with the date, GOES class, position, and peak time are given in Table \ref{table1}. Steps of our analysis are explained in the following subsections. \\

\subsection{Time profiles}
\begin{figure*}
\centering
\includegraphics[scale=0.85]{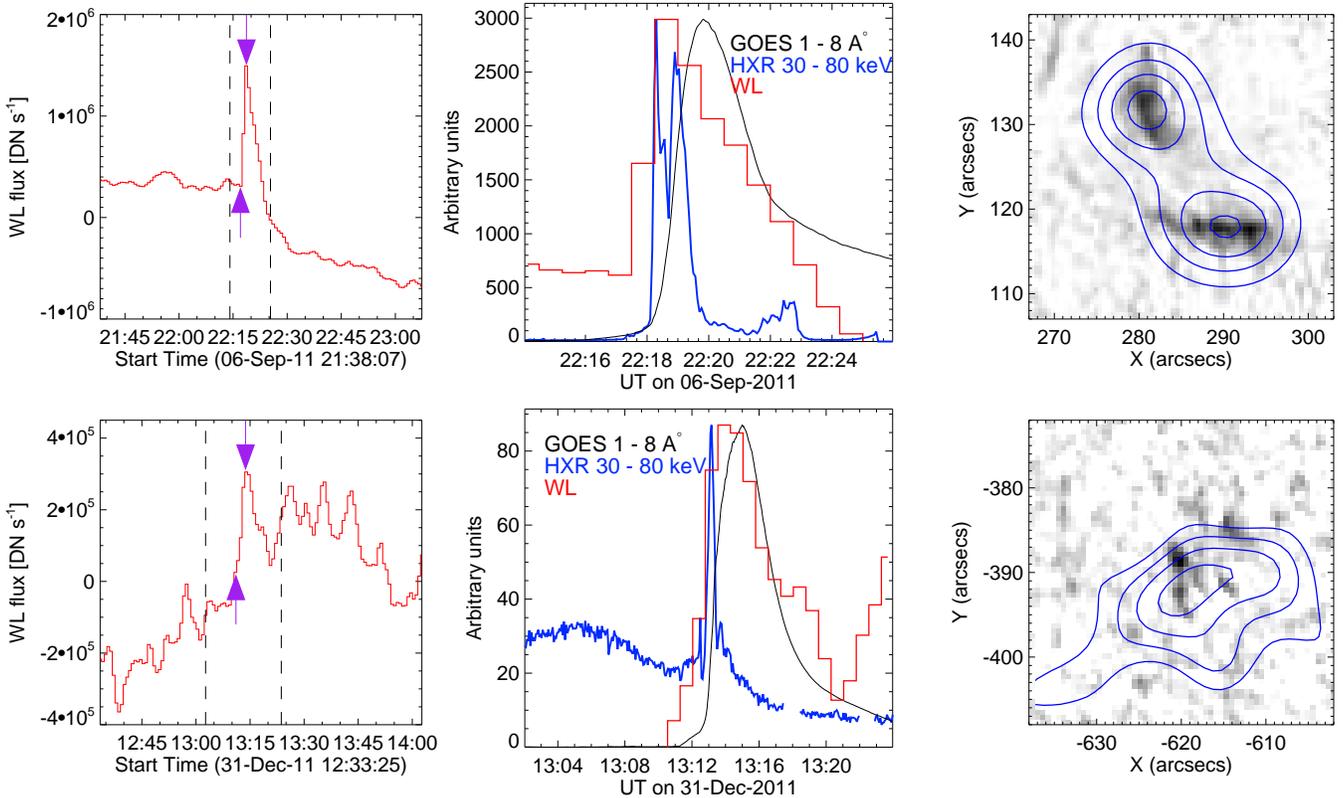}
\caption{Time profiles and images of an intense (SOL2011-09-06 (X2.1)) and a weak (SOL2011-12-31 (M2.4)) event in our sample: Left panels show the WL emission (calculated inside the 30\% countour of HXR emission) time profiles for a long time span around the time of peak emission to compare the increase due to the flare relative to the fluctuations of the non-flaring active region. The vertical lines indicate  time ranges used in the plots in the central panels, while the purple arrows indicate peak- and pre-flare frames used for the analysis. Central plots show WL flux in red, HXR flux at $30-80$ keV in blue and GOES flux at $1-8$ \AA \hspace{1pt}  in black. Right panels show pre-flare subtracted images of events in HMI with RHESSI contours of 30\%, 50\%, 70\% and 90\% of the maximum emission in the $30-80$ keV CLEAN-image overplotted in blue.  }
\label{fig1}
\end{figure*}

First, we plotted the WL lightcurve (calculated by summing the WL emission over the HXR footpoints) for a time span of $\sim$1.5 hours around peak time, in order to see the evolution of the non-flaring Sun relative to the enhancement of WL emission during solar flares (see left panels in Figure \ref{fig1}). For making the images as shown in Figure \ref{fig1} and calculating WL fluxes for later analysis, we subtracted a pre-flare image  from the peak image. For each flare, we chose the pre-flare image as the one closest to the average value (of all frames) inside the time range of a few minutes before the flare peak time (using images far from the peak time is not applicable, as the flaring region can evolve substantially on larger time scales). Peak- and pre-flare frames are indicated in Figure \ref{fig1} with purple arrows. From the non-flaring temporal variation we infer a conservative (upper limit) error estimate as the maximum of this variation around peak time, and we give this value in Table \ref{table1}.  Vertical lines indicate the peak time range, for which we plot GOES, HXR and WL time profiles (middle panels in Figure \ref{fig1}).\par
We note here that the detection sensitivity for WL emission is significantly enhanced when considering time profiles summed over the flaring region (such as shown in Figure \ref{fig1}) in combination with inspecting running difference images by eye.

\subsection{HXR imaging and WL flux derivation}
 For each flare we made an image in the HXR range at $30-80$ keV using detectors 3$-$6 and the CLEAN algorithm \citep{Hurford02}, and overplotted it on the WL pre-flare subtracted image. Two examples are shown in the right panels of Figure 1. We identified a systematic offset between WL and HXR footpoints. For each flare, we computed polar angles of the maximum HXR and WL emissions, and  the difference between them. The histogram distribution of polar angle differences peaks at 0.15 degrees. This difference is most noticeable at the solar limb,  where it translates to a spatial separation of  $\sim2.5''$. We speculate that this difference comes from an error in the roll-angle calibration in one of the two instruments. For our study, we overcome this problem by calculating the WL flux inside the 30\% contour of maximum HXR emission. This is a large enough area so that most  WL enhancement is included, despite the offset in position, for all flares.

 \subsection{HXR spectral fitting} 
 We used the OSPEX package for the calculation of  HXR fluxes and  spectral indices. We performed spectral fitting above 18 keV  with the integration time of 45 seconds around the time of maximum WL emission, in order to match the integration time of HMI. We fitted a thermal + power-law model to the data. Fluxes at higher energies were extrapolated from the fluxes at 30 keV and the spectral indices using the standard formula for HXR power spectrum:
\begin{equation}
F(E)=F(30) \cdot(E/30)^{-\gamma},
\label{eq1}
\end{equation}
where $F(E)$ stands for photon flux at energy $E$, $F(30)$ is the photon flux at 30 keV, and $\gamma$ is the spectral index. Because of the steepness of the HXR spectrum, there are many more photons with lower energies. The energy of 30 keV is chosen because it is the lowest energy that contains a negligible amount of thermal emission, and still contains large fluxes of non-thermal photons. As we are mainly interested in the spectral slope around 30 keV, we only fit a thermal + single power-law, without a break. If these values are used for the extrapolation to higher energies, it should be considered that hard X-ray flare spectra generally have breaks \citep[e.g., ][]{Dulk72}, and our simplified approach could lead to the overestimation of HXR fluxes at higher energies, in particular above $\sim100$ keV. Using the thick target approximation \citep[e.g., ][]{Brown71}, the energy deposition by non-thermal electrons can be derived from the HXR spectral parameters for a given low energy cut-off \citep[e.g., ][]{Saint-Hilaire05}.

\section{Results}\label{section3}

\subsection{The data analysis steps for two examples}
In Figure \ref{fig1} we present lightcurves and images of two flares, an intense event in which a clear temporal and spatial correlation between WL and HXR fluxes can be seen, and one of the weakest events in our sample. The SOL2011-09-06 (X2.1) event is an example with very good HXR counting statistics and a WL source well above the pre-flare emission. The post-flare WL emission for this event does not recover to the pre-flare value, which appears to be due to the flare-related permanent change in the active region. For the few events with similarly good statistics, the observations are good enough to compare details of the flare ribbons (e.g. Krucker et al. 2011). However, such a study is not the focus of this statistical work where we try to include a large number of events. On the other hand, the SOL2011-12-31 (M2.4) event has rather poor statistics and a low contrast and the details of the source morphologies cannot be compared. Nevertheless, the WL flux integrated over the HXR source has a local maximum roughly at the expected time (the peak emission in WL for this event is $\sim45$ seconds after peak time of HXR emission). Post-flare time profile of this event shows a few additional peaks. Due to its weak WL emission (an order of magnitude weaker than the WL emission of SOL2011-09-06 (X2.1)), these sub-peaks most probably represent just the usual background fluctuations of the active region rather than the subsequent peaks of WL emission due to the flare. We included this event in our sample although with a large error bar at $\sim$80\% of the observed value (see Table 1). 

Figure 1 clearly shows that careful examination of lightcurves is essential for determining the appropriate pre-flare image and for imaging the flare-enhanced emission in WL. We also point out co-temporality between HXR and WL emission in the majority of our events, within our 45 second cadence. A higher cadence of a few seconds would be needed for a closer inspection of WL$-$HXR co-temporality. WL flux for most events has a longer decay phase, which usually  lasts a few minutes. Generally, the good correlation of HXR and WL time profiles, with a longer decay phase in WL, is in agreement with previous observations \citep[e.g.,][]{Hudson92, Xu06, Matthews03}. \par

\subsection{Relation between HXR and WL fluxes}
\subsubsection{The limb darkening correction}
Depending on the height of the WL source above the photosphere, we expect to see a limb darkening effect (i.e. stronger absorption for events closer to the limb due to the enhanced column density along the line-of-sight). For a WL source originating in the photosphere, the classical limb darkening function could be a good approximation. For sources at higher altitudes, the absorption along the line of sight could be significantly less than the photospheric value. Additionally, the roughness (i.e. deviation from spherical symmetry) will further influence the overall limb darkening effect. \par
\begin{figure}[hbtp]
\centering
\includegraphics[scale=0.4]{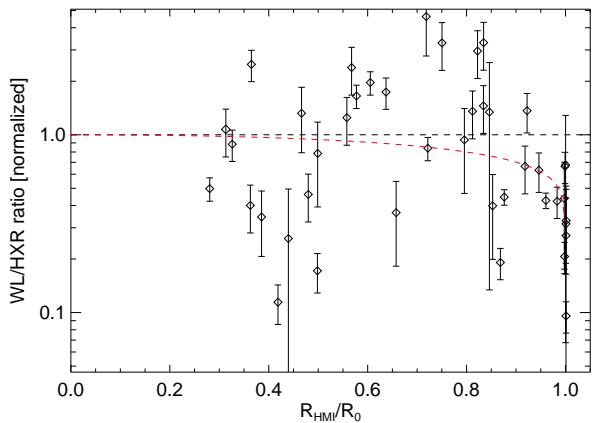}
\caption{Plot of the ratio of WL to HXR fluxes vs. radial distance of WL emission from Sun center. $R_{HMI}$ and $R_0$ denote the distance  of the maximum WL emission from Sun center and Sun's radius at flaring time, respectively. Ratios of WL and HXR fluxes are normalized to the average ratio for  flares with radial distance $<700$ arcsec. The photospheric limb darkening function is over-plotted in red.}
\label{fig2}
\end{figure}

In Figure \ref{fig2} we plot the non-corrected  WL/HXR flux ratios vs. the radial distance from Sun center of the WL emission. Since we do not know if all flares have a constant WL/HXR ratio (some flares might be more efficient in producing WL emission than others), fitting a limb darkening function directly to these values is not applicable. Instead, we use the average ratio of disk flares (radial distance $<700$ arcsec) as a reference for events that are not affected by limb darkening. This value is represented by a horizontal line in the plot. The red curve is the limb darkening function taken from \cite{Pierce77} for the wavelength closest to 617.3 nm. While no firm conclusions can be made, it can be seen that all flares near the limb (radial distance $>900$ arcsec) have substantially smaller WL fluxes than the average. Our conclusion is that applying the limb darkening correction is better than applying no correction at all, even though it may not be perfect. \par
In the above analysis of the limb darkening of WL/HXR ratios, we implicitly assumed that  center-to-limb variation in HXR at 30 keV is negligible, which is in agreement with the studies of \cite{Datlowe77} and \cite{Kasparova07}.

\subsubsection{Fitting of the data and the correlation coefficients}

\begin{figure*}[hbtp]
\centering
\includegraphics[scale=0.9]{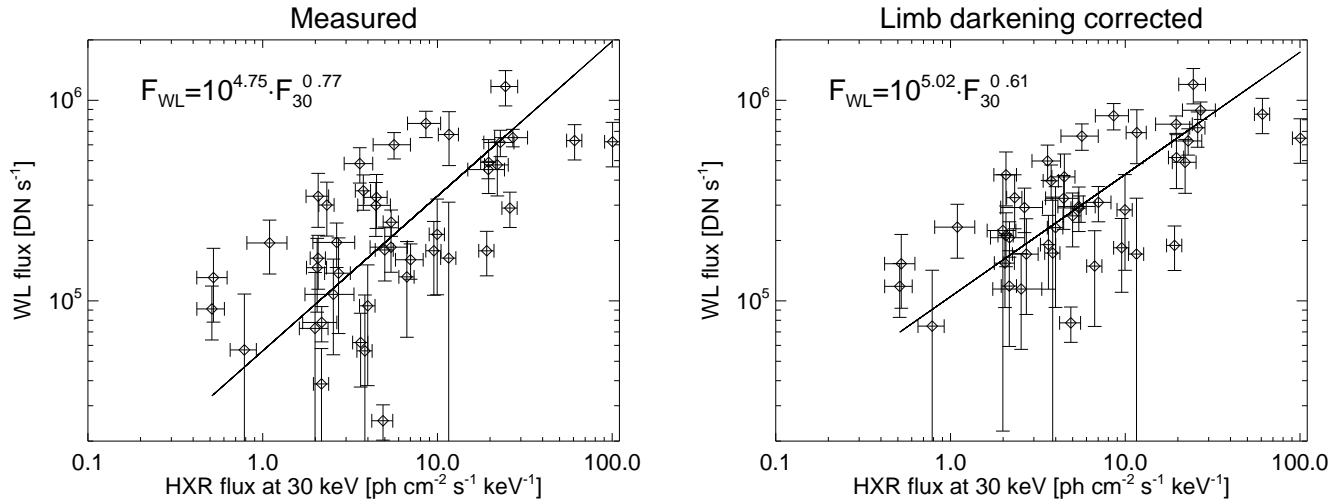}
\caption{Left: Correlation between measured WL fluxes and HXR fluxes for 43 events analyzed in our study. Right: Same as on the left, but with limb-darkening-corrected WL fluxes.}
\label{fig3}
\end{figure*}

As described in Section \ref{section2}, fluxes in HXR were calculated for $E=30$ keV via spectral fitting, while WL fluxes were calculated inside the 30\% RHESSI contour from pre-flare subtracted images. When applying spectral fitting, we used different detectors for each flare, depending on the detector state of health that varies during the large time span of  4 years considered in this survey. The HXR flux of each flare is the average of the fluxes given by the `healthy' detectors. The error bar is estimated as the standard deviation of the HXR fluxes given by these detectors around the average value. Error bars in WL are given by the maximum emission of the fluctuation of the background  $\sim15$ minutes before a flare occurred.  These are conservative (upper limit) values, especially for fainter flares. In  Figure \ref{fig3}, we present the correlation between HXR and WL fluxes for two cases: measured WL fluxes, and WL fluxes with the limb darkening correction applied. Since the calculated fluxes span two orders of magnitude, we present our results in a log$-$log plot.

We use a fit to both sets of values in the form $F_{\rm{WL}}=A\cdot(F_{\rm{30}})^{b} $, where $F_{\rm{WL}}$ is white light flux, $F_{\rm{30}}$ is HXR photon flux at 30 keV, $A$ is a constant, and $b$ is the power law index. Since we \textit{a priori} do not know the mutual dependence of the WL and HXR fluxes, we use a bisector regression method \citep{Isobe90}. For the power law index we get the values of $b=(0.77 \pm 0.10)$ and $b=(0.61 \pm 0.07)$ for the observed and limb darkening corrected values, respectively.  The correlation coefficient has values of 0.62  and 0.68  at the energy of 30 keV. \par

\begin{figure}[hbtp]
\centering
\includegraphics[scale=0.4]{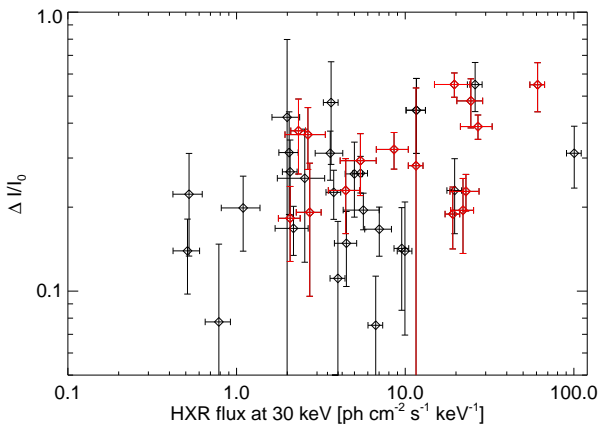}
\caption{Plot of the relative enhancement of the WL emission vs. the HXR flux. The correlation is worse than for the absolute enhancement, with the correlation coefficient of $0.36$. Red points denote the best events in the sample, and for this subset the correlation coefficient reaches the value of 0.4.}
\label{fig4}
\end{figure}

Another frequently used parameter for quantifying WL emission in solar flares is the relative enhancement of the WL emission (defined as $\Delta I/I_0=(I-I_0)/I_0$, $I$ and $I_0$ being the peak- and pre-flare fluxes, respectively), as it compares the pre-flare photospheric flux to the flare enhancement. In order to study its connection to the HXR emission, we made pre-flare subtracted images and normalized them to the pre-flare images, in order to get the information on the relative enhancement of each individual pixel, for each flare. We chose the brightest pixel in the relative enhancement maps as our estimate of the relative enhancement of the WL emission. Using the average value of relative enhancements over the area of the 30\% HXR contour would lead to underestimation of the relative enhancement, since this area contains many non-flaring pixels. Here we note that we only analyzed on-the-disk flares, as otherwise $I_0$ does not come from the underlying photosphere (these values are also not included in Table \ref{table1}). The plot of maximum relative enhancements vs. HXR fluxes is presented in Figure \ref{fig4}. Single pixel values of $\Delta I/I_0$ span the range $0.1-0.6$. The correlation is weaker than in the former case of absolute enhancements, with the correlation coefficient being 0.36. The correlation is only slightly improved if only the best events from our sample are chosen (red points in the figure), when the correlation coefficient reaches the value of 0.4. Our results slightly favor the absolute enhancement over the relative enhancement as the more relevant quantity for the WL formation; however, further studies must be made in order to test this claim in more detail. \par

\begin{figure}
\centering
\includegraphics[scale=0.4]{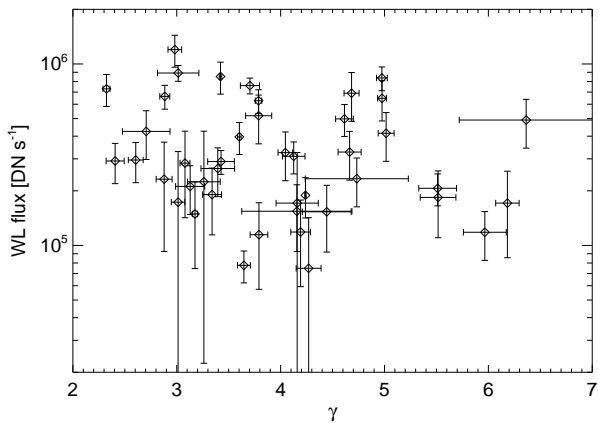}
\caption{Plot of WL fluxes (limb-darkening correction applied) vs. the power law indices of HXR spectra  for 43 events analyzed in our study. There is no correlation between the two quantities.}
\label{fig5}
\end{figure}

We also analyzed the relation between spectral indices and WL fluxes. As can be seen in Figure \ref{fig5}, there is no correlation between the two quantities. This result suggests that the highest energy electrons play a minor role in the production of WL emission, which makes sense from an energetical point of view as the low-energy electrons carry more energy than the high-energy electrons (due to their larger numbers).

\subsubsection{Above the limb flares}
Of particular interest are the three events with emissions above the solar limb reported by Krucker et al. (2015) that are also included in the survey presented here. The HXR and WL emissions for these events occur within one degree of limb passage and the observed radial position therefore directly  translates into actual altitude (within the measurement uncertainties). The observations show co-spatial WL and HXR source peaking at altitudes above the photosphere of around 800 km. The WL to HXR ratios for these events are between 0.095 to 0.328 if normalized to the averaged on-disk ratio. This indicates a strong limb darkening effect, but it additionally could be the case that emission from lower altitudes is completely absorbed and only the top part of the source is observed. Our statistical survey alone cannot distinguish the two explanations. To look into this issue, a detailed theoretical study of the optical depth as a function of height above the photosphere needs to be performed, but this is outside the scope of this observational paper.

\subsection{The correlation between HXR energies and WL fluxes}
\begin{figure*}[hbtp]
\centering
\includegraphics[scale=0.9]{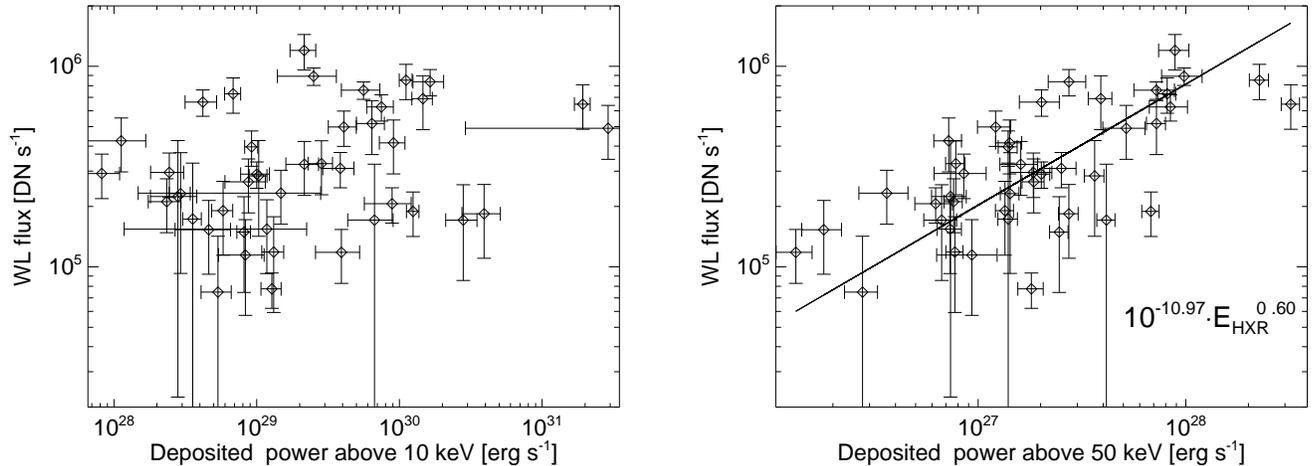}
\caption{\textit{Left}: The limb-darkening-corrected WL flux vs. the power deposited by electrons above 10 keV. Since there is no clear correlation in this case, the fit to the values is not shown. \textit{Right}: Same as on the left, but for electrons above 50 keV.}
\label{fig6}
\end{figure*}

The correlations presented so far are all with directly observed quantities. From a physical point of view, it is desirable to directly compare the energy input (i.e. energy deposition by non-thermal electrons) vs. energy output (i.e. radiative losses in the optical range). However, such estimates rely on model assumptions. From the HXR spectra we can use the classical thick target assumption to derive the total energy deposition of flare-accelerated electrons above a given electron energy \citep[e.g., ][]{Brown71}. Since we only have single-frequency WL observations, it is extremely difficult to estimate the total radiative losses in WL.  For this work we are therefore limited to use the observed WL fluxes to compare with the energy deposition by non-thermal electrons.  Nevertheless, this allows us to find the energy range of electrons that best correlates with the observed WL fluxes.  We calculated the deposited power by electrons for cut-off energies ranging from 10 to 100 keV. In Figure \ref{fig6} we present scatter plots for two cut-off energies, 10 and 50 keV, respectively.  The best correlation is found for the deposited power by non-thermal electrons above 50 keV, while the correlation decreases rapidly if the power provided by lower energy electrons is added as well. The slopes in the scatter plots of WL flux vs. deposited power are $\sim0.6$ in the $40-70$ keV range, similar to the slopes in the fluxes scatter plots. \par

\begin{figure}[hbtp]
\centering
\includegraphics[scale=0.4]{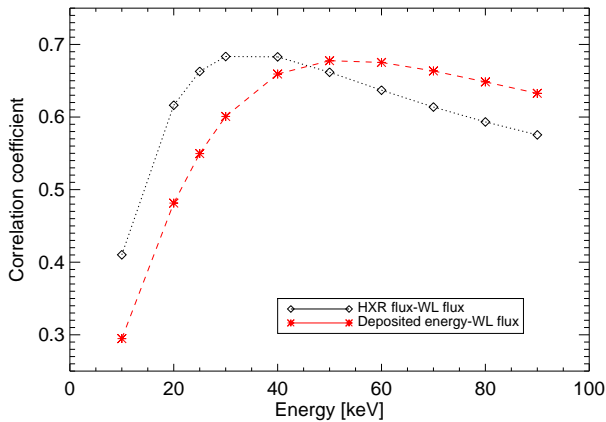}
\caption{Correlation coefficient between the logarithmic values of HXR fluxes at different energies and WL fluxes (diamonds), and between the logarithmic values of the deposited energy by non-thermal  HXR electrons above different threshold energies and WL fluxes (stars). The correlation coefficient decreases steeply for energies below 30 keV. It has the maximum for HXR fluxes at 30 keV, and for HXR energies for a cut-off energy around 50 keV.}
\label{fig7}
\end{figure}

Figure \ref{fig7} shows the behavior of the correlation coefficient between  WL fluxes and power deposited by non-thermal electrons above different cut-off energies in one case, and between  WL fluxes and HXR photon fluxes at different energies in the other case. Both show similar behavior, steep decrease for energies below 30 keV and a gradual decrease towards higher energies. This behavior can be explained within the standard thick-target model as the lowest energy electrons are stopped too high in the solar atmosphere to be responsible for WL production. The highest energy electrons, on the other hand, do not carry enough energy to account for the overall WL emission. Obvious is also a difference in the peak energy of the two curves, which is at 30 keV for the flux$-$flux correlation coefficient, and at 50 keV for the power$-$flux correlation coefficient. This difference is a direct result of the thick-target assumption. As electrons at a given energy produce photons at all lower energies, it is expected that the correlation coefficient peaks at a higher value for the electron energy than for the photon energy. The difference we get is in  agreement with the study of \cite{Saint-Hilaire05}, where they calculated the ratio of the turnover energy in the photon spectrum and the cut-off energy in the electron beam distribution to be $\sim0.6$. Hence, electrons with energy of 50 keV typically produce  photons of 30 keV, which is in agreement with the curves shown in Figure \ref{fig7}. \par

\section{Discussion and conclusions}\label{section4}

Simultaneous observations of RHESSI and HMI for 43 flares spanning GOES classes M and X have provided valuable information about the correlations in time, space and intensity of HXR and WL fluxes in solar flares. Most of the previous observations of WL flares have focused on individual events. In this study, we present statistical results with the following conclusions.\par

\textbf{Temporal:} With the rather low cadence of HMI standard data products used in this study, the conclusions are limited. Nevertheless, the peaks of WL emission are co-temporal with HXR emission for most events, at least within the time resolution of 45 seconds for WL emission. The WL emission has a longer decay phase, typically lasting a few minutes, as was previously reported \citep[e.g.,][]{Hudson92, Xu06, Matthews03}.\par

\textbf{Occurrence:} Our statistical survey does not mandate the existence of HXR flares without WL emission. All large flares (above GOES class M5) show WL counterparts. The lack of WL emission in smaller flares is rather due to the difficulties in distinguishing the WL flare signal from the general time variation of the non-flaring regions at optical wavelengths than the actual absence of WL flare emission. As our study does not include microflares (i.e. GOES C class and smaller), we cannot state anything regarding this group of events. Nevertheless, WL emission has been reported from C-class flares \citep{Jess08} and there are GOES A- and B-class flares with relatively intense HXR emissions (order of 0.1 photon/s/cm$^2$/keV, Ishikawa et al. 2011) for which WL emission could potentially be detectable.\par

\textbf{Intensity:} There is a clear correlation between WL and HXR fluxes. However, the correlation is not linear, but better described by a power law with an index of $(0.77 \pm 0.10)$ and $(0.61  \pm 0.07)$ for measured and limb darkening corrected WL fluxes, respectively.  
The absence of a linear correlation is not surprising, as the conversion of HXR flux and WL flux at the HMI wavelength to actual energy input and output does not need to be linear.
The correlation coefficient between logarithmic values of WL and HXR fluxes is 0.68 for the energy of 30 keV. The scatter of the observed and limb darkening corrected values is relatively large with extreme values being up to a factor of $\sim7.5$ and $\sim2.5$ different from the fitted curve for the observed and limb darkening corrected values, respectively. The WL flux correlates best with the HXR flux around 30 keV with a  decrease in the correlation coefficient for lower and higher (non-thermal) photon energies. The correlation coefficient decreases if the relative enhancements are used to describe the WL emission, and it has a value of $0.36$ at the energy of 30 keV, suggesting that the pre-flare WL emission is not a key factor in the production of WL emission. \par

\textbf{Spectral:}  There is no correlation between  HXR spectral indices in the $30-100$ keV range and WL fluxes. This suggests that high energy electrons play a minor role in the production of WL emission. \par

\textbf{Energy deposition:} There is a clear correlation between the deposited power by flare-accelerated electrons and WL fluxes best seen for electron energies in the range $40-70$ keV with a maximum correlation coefficient of 0.68 at 50 keV. The correlation coefficient decreases gradually towards higher energies, while the steep decrease is observed for lower energies. This suggests that the lowest energy electrons are not involved in the production of the WL emission. The most likely explanation for this behavior is that low-energy electrons accelerated in the corona are stopped too high in the atmosphere to play a role in the WL formation.

Our results are consistent with earlier findings from a similar survey of Yohkoh HXR and WL data \citep{Matthews03}. The correlation in flux, however, is a new result. The existence of a correlation does not contradict the work by Matthews et al. 2003 and might be due to our larger sample size in combination with the enhanced sensitivity of HMI and RHESSI compared to Yohkoh. Similarly, the observed HXR spectral indices are much more accurate and the absence of a correlation is now a  solid result.\par

In summary, we conclude that WL and HXR fluxes show a clear correspondence in space, time, and intensity.  The newly found correlation between energy deposition and WL flux is a further indication that  flare-accelerated electrons are the main contributor to the WL formation. The absence of detectable WL emission in weaker flares is most probably caused by the weak contrast of WL flare emission when compared to non-flaring Sun temporal variations, and it is unlikely that a second class of flares exists without WL emission, at least for larger flares. As smaller flares are expected to also have fainter WL emissions, their detection is even more difficult, making it unlikely to get conclusive observations of flares without WL counterparts. \par

As a possible explanation for the newly found correlation between energy deposition by $>$50 keV electrons and WL formation, we would like to put forward the following scenario: If HXR and WL sources are indeed co-spatial for all flares as suggested by observations of a few single events \citep[]{MartinezOliveros12, Battaglia12, Krucker15},  the accelerated electrons above 50 keV produce the WL emission by directly heating dense layers by collisions to moderate temperatures ($\sim 10^4$ K). The low-energy end of the accelerated electrons lose their energy above the  WL/HXR source at lower densities producing hot plasma ($\sim 10^7$ K) in the flare ribbons that is radiating at EUV and SXR wavelengths \citep[e.g.,][]{Hudson94, Graham13}. Through radiative cooling, the WL source is dissipating the part of the flare energy that was previously stored in the high-energy ($>50$ keV) tail of the accelerated electrons. The energy in the low-energy electrons goes into heating of the plasma, part of which then evaporates and produces the main flare loop.  An open question  is the observation of rather low altitudes of the WL/HXR sources \citep{MartinezOliveros12}, which is challenging to explain within the standard thick-target model. In any case, our findings alone do not exclude a back-warming model where the WL source originates from deeper layers than the HXR source. \par

The next step in the research will be the investigation of the relation between the energy deposited by non-thermal electrons and the energy contained in WL for different threshold HXR energies. This requires coverage of the optical range by observations at several wavelengths, which are only available for a few well-observed events \citep[e.g.,][]{Milligan14}. To test our hypothesis that the low-energy electrons produce hot plasma at higher altitudes, we are planning to analyze SDO/AIA data for the sample of events presented here.  By applying  the newly available de-saturation algorithm for AIA images \citep{Schwartz14}, we will be able to determine EUV data points for many flares in our current statistics. This will allow us to derive the thermal energy content in the flare ribbon and compare it to the deposited energy by non-thermal electrons.

\acknowledgments

The work was supported by Swiss National Science Foundation (200021-140308), by the European Commission through HESPE (FP7-SPACE-2010-263086), and through NASA contract NAS 5-98033 for RHESSI at UC Berkeley.

\bibliographystyle{apj}
\bibliography{journals,whitelight}

\newpage

\hspace*{-8.0cm}\begin{deluxetable}{cccccccccc}
\tabletypesize{\footnotesize}
\tablecolumns{10} 
\tablewidth{0pt} 
\tablecaption{List of 43 white light flares analyzed}
\tablehead{
\colhead{Date} & 
\colhead{GOES}   & 
\colhead{Position}    & 
\colhead{Peak time}  & 
\colhead{$F_{\textrm{30keV}}$}   & 
\colhead{Limb}  & 
\colhead{$F_\textrm{WL}$(corr.)}    & 
\colhead{$\Delta I/ I_0$}    & 
\colhead{Spectral} &
\colhead{Deposited power}\\
\colhead{} & 
\colhead{class} & 
\colhead{[arcsec]} & 
\colhead{[UT]} & 
\colhead{[ph cm$^{-2}$ s$^{-1}$ keV$^{-1}]$} & 
\colhead{factor} &
\colhead{[10$^5$ DN s$^{-1}$]} & 
\colhead{} & 
\colhead{index}   &
\colhead{[$10^{27}$ erg s$^{-1}$]}  }
\startdata
\centering
2011 Feb 15 & X2.2 & [210, -220] & 01:53:42 & 11.6 $\pm$ 1.5&1.02 & 6.9 $\pm$ 2.1&0.45$\pm$0.13 & 4.68 $\pm$ 0.07 &3.90 $\pm$ 0.51\\
2011 Feb 18 & M1.4 & [765, -275] & 13:01:57 & 0.51 $\pm$ 0.09&1.30 & 1.2 $\pm$ 0.4&0.14$\pm$0.04 & 5.97 $\pm$ 0.21&0.13 $\pm$ 0.03\\
2011 Feb 18 & M6.6 & [755, -270] & 10:10:57 & 2.07 $\pm$ 0.31& 1.28& 4.2 $\pm$ 1.1&0.18$\pm$0.05 & 2.71 $\pm$ 0.23&0.72 $\pm$ 0.11\\
2011 Feb 24 & M3.5 & [-925, 275] & 07:31:13 & 3.99 $\pm$ 0.41 &2.45 &2.5 $\pm$ 1.5 &0.11$\pm$0.07 & 2.88 $\pm$ 0.08&1.43 $\pm$ 0.15\\
2011 Mar 07&M3.7 & [615, 560]& 20:01:14&9.96 $\pm$ 0.35& 1.33&2.8 $\pm$ 1.4&0.14$\pm$0.07 &3.08 $\pm$ 0.05&3.64 $\pm$ 0.39\\
2011 Mar 09 & X1.5 & [190, 275] & 23:20:44 & 22.0 $\pm$ 3.4 &1.03& 4.9 $\pm$ 2.5 &0.19 $\pm$0.06 & 6.36 $\pm$ 0.64&5.17 $\pm$ 1.18\\
2011 Mar 14 & M4.2 & [705, 340]& 19:51:30 & 4.48 $\pm$ 0.68 & 1.27&4.2 $\pm$ 1.7&0.15 $\pm$0.04&5.02 $\pm$ 0.08&1.42 $\pm$ 0.20\\
2011 Mar 15& M1.0& [750, 325] & 00:21:30& 0.79 $\pm$0.13 & 1.31&0.75 $\pm$ 0.67 &0.08$\pm$0.07 &4.27 $\pm$ 0.12 &0.28 $\pm$ 0.05\\
2011 Jul 30 & M9.3 & [-525, 170] & 02:08:11 & 8.59 $\pm$ 1.84&1.09& 8.4 $\pm$ 1.3&0.32$\pm$0.05 &4.97 $\pm$ 0.05 &2.74 $\pm$ 0.56\\
2011 Sep 06 & X2.1 & [280, 130] & 21:18:37 & 24.5 $\pm$ 4.3 &1.02 &12.0 $\pm$ 2.4&0.48$\pm$ 0.10&2.98 $\pm$ 0.07 &8.87 $\pm$ 1.47\\
2011 Sep 24 & M5.8 & [-740, 155]& 20:34:20 & 2.72 $\pm$ 0.46&1.25& 1.7 $\pm$ 0.8&0.19$\pm$0.10 &6.18 $\pm$ 0.11 &0.67 $\pm$ 0.12\\
2011 Sep 24 & X1.9 & [-815, 165] &09:36:35 & 27.0 $\pm$ 5.8 &1.37 &8.9 $\pm$ 0.9 &0.39 $\pm$0.04&3.01 $\pm$ 0.20 &9.81 $\pm$ 2.17\\
2011 Sep 26 & M4.0 & [-520, 120]& 05:06:35 & 4.44 $\pm$ 0.94 & 1.08&3.2 $\pm$ 1.0&0.23 $\pm$0.07&4.05 $\pm$ 0.07 &1.60 $\pm$ 0.33\\
2011 Dec 26 & M2.3 & [635, 325]& 20:16:55 & 0.52 $\pm$ 0.10& 1.17&1.5 $\pm$ 0.6 &0.22$\pm$0.09 &4.44 $\pm$ 0.23 &0.18 $\pm$ 0.04\\
2011 Dec 31& M2.4 & [-620, -395]& 13:13:10& 1.09 $\pm$ 0.28&1.20 &2.3 $\pm$ 1.9&0.20$\pm$ 0.06&4.73 $\pm$ 0.50 &0.36 $\pm$ 0.10\\
2012 Mar 09 & M6.3 & [50, 380] & 03:40:15 & 9.53 $\pm$ 0.95 & 1.03&1.8 $\pm$ 0.9&0.14 $\pm$0.06&5.52$\pm$ 0.17 &2.73 $\pm$ 0.30\\
2012 May 10& M5.7& [-385,255]& 04:16:23&19.63 $\pm$ 0.86& 1.06&5.2 $\pm$ 1.6& 0.23$\pm$0.07& 3.78 $\pm$ 0.13 &7.22 $\pm$ 0.76\\
2012 Jun 03&  M3.3 &[-565, 275] & 17:53:10 & 6.68 $\pm$ 0.67 &1.13 &1.5 $\pm$ 0.7 & 0.8$\pm$0.04&3.18 $\pm$ 0.02 &2.46 $\pm$ 0.25\\
2012 Jul 04& M5.3&[290, -340]  &09:54:40 & 2.05 $\pm$ 0.27&1.05 &1.5 $\pm$ 0.6& 0.31$\pm$0.13 &4.16 $\pm$ 0.53 & 0.73 $\pm$ 0.10\\
2012 Jul 05 & M4.7 & [415, -335] & 03:35:10 & 2.33 $\pm$ 0.23&1.09 &3.3 $\pm$ 0.7 & 0.38$\pm$0.11& 4.66 $\pm$ 0.11 &0.78 $\pm$ 0.08\\
2012  Jul 05& M6.1& [500, -340]& 11:44:10& 3.77 $\pm$ 0.38 & 1.12&4.0 $\pm$ 0.8 &0.23$\pm$0.05 & 3.60 $\pm$ 0.02 &1.40 $\pm$ 0.15\\
2012 Jul 06 & M2.9 & [590, -330] & 01:38:55& 5.43 $\pm$ 0.54&1.18 &2.9 $\pm$ 0.3&0.26$\pm$0.04 &3.43 $\pm$ 0.13 &2.01 $\pm$ 0.20\\
2012 Jul 19 & M7.7 & [925, -200]& 05:21:40 & 2.17 $\pm$ 0.22 & 3.08&1.2 $\pm$ 0.6 & $-$&4.19 $\pm$ 0.09 &0.77 $\pm$ 0.07\\
2012 Aug 06 & M1.6 &[-915, -230] & 04:35:54& 1.99 $\pm$ 0.37&3.08 &2.2 $\pm$ 1.8 &0.42$\pm$0.38 &3.26 $\pm$ 0.16 &0.74 $\pm$ 0.14\\
2012 Oct 23 & X1.8 & [-800, -260] & 03:15:30 &60.9 $\pm$ 6.1& 1.35&8.5 $\pm$ 2.1& 0.55$\pm$0.11&3.42 $\pm$ 0.01 &22.59  $\pm$ 2.38\\
2012 Nov 20 & M1.7& [950, 200]& 12:39:26 &3.84 $\pm$ 0.38&3.08 &1.7 $\pm$ 1.5& $-$ &3.01 $\pm$ 0.07 &1.40 $\pm$ 0.15\\
2013 May 13 & X1.7 & [-930, 200]& 02:09:37 &4.88 $\pm$ 0.67&3.08 &0.8 $\pm$ 0.4& $-$&3.65 $\pm$ 0.06 &1.80 $\pm$ 0.25\\
2013 May 13 & X2.8&[-925, 180] & 16:03:37 & 26.0 $\pm$ 2.6& 2.51&7.3 $\pm$ 1.5 & 0.55$\pm$0.11&2.32 $\pm$ 0.03  &8.13 $\pm$ 0.81\\
2013 May 15 & X1.2 & [-850, 200] &01:41:53 & 5.00 $\pm$ 0.59 & 1.48&2.7 $\pm$ 0.8&0.26$\pm$0.08 & 3.39 $\pm$ 0.16 &1.85 $\pm$ 0.23\\
2013 Oct 25 & X1.7 & [-910, -160] & 07:58:14 & 19.6 $\pm$ 4.6&1.68&7.6 $\pm$ 0.8& 0.55$\pm$0.06& 3.70 $\pm$ 0.09 &7.22 $\pm$ 1.59\\
2013 Oct 28 & M5.1 &[-440, -195] & 15:10:59 &2.54 $\pm$ 0.79&1.06&1.1 $\pm$ 0.6&0.25$\pm$0.13 & 3.79 $\pm$ 0.08 &0.93 $\pm$ 0.30\\
2013 Oct 28 & X1.0 & [910, 40]& 01:59:44 &5.42 $\pm$ 1.31&1.60&3.0 $\pm$ 0.7&0.29$\pm$ 0.07&2.61 $\pm$ 0.07 &1.84 $\pm$ 0.42\\
2013 Nov 10&X1.1 &[230, -285] &05:13:12 &3.59 $\pm$ 0.67&1.03&5.0 $\pm$ 1.0&0.31 $\pm$0.06&4.61 $\pm$ 0.09 &1.21 $\pm$ 0.22\\
2014 Jan 07 & X1.2&[-220, -170] & 10:11:39 & 22.9 $\pm$ 4.5 &1.02&6.3 $\pm$ 1.6&0.23$\pm$0.03 & 3.79 $\pm$ 0.03 &8.43 $\pm$ 1.76\\
2014 Jan 27 &M4.9 & [-940, -260]& 22:09:24&2.17 $\pm$ 0.48 &2.64& 2.1 $\pm$ 0.4&0.17 $\pm$0.03&5.51 $\pm$ 0.18 & 0.62 $\pm$ 0.13\\
2014 Feb 07 &M1.9 & [765, 265]&10:28:10 &2.08 $\pm$ 0.09& 1.30&2.1 $\pm$ 0.6&0.27$\pm$0.08 & 3.13 $\pm$ 0.14  &0.76 $\pm$ 0.08\\
2014 Mar 12 &M9.3 &[910, 270] &22:31:59 &7.02 $\pm$ 1.27&1.93 &3.1 $\pm$ 0.6&0.17 $\pm$0.03& 4.13 $\pm$ 0.11 &2.52 $\pm$ 0.44\\
2014 Mar 29& X1.0 & [515, 265] &17:46:16 &5.65 $\pm$ 1.36& 1.10&6.6 $\pm$ 0.7&0.20$\pm$0.03 &2.88 $\pm$ 0.05 &2.02 $\pm$ 0.44\\
2014 Jun 11 & X2.2 & [-820, -305] & 09:05:10 &2.65 $\pm$ 0.71& 1.49&2.9 $\pm$ 0.6&0.36 $\pm$0.09&2.41 $\pm$ 0.09 &0.85 $\pm$ 0.24\\
2014 Oct 16 &  M4.3& [-935, -225] &13:02:00 &3.63 $\pm$ 0.36& 3.08&1.9 $\pm$ 0.8&0.47$\pm$0.19 &3.34 $\pm$ 0.09 & 1.35 $\pm$ 0.13\\
2014 Oct 22 &M8.7 & [-390, -295] & 01:38:45 &19.1 $\pm$ 1.9& 1.06&1.9 $\pm$ 0.3&0.19$\pm$ 0.05&4.24 $\pm$ 0.01 & 6.79 $\pm$ 0.65\\
2014 Oct 22 & X1.1& [-170, -320]&14:06:30 &100.5 $\pm$ 4.4& 1.04&6.5 $\pm$ 1.6&0.31 $\pm$0.08&4.98 $\pm$ 0.04 &32.01 $\pm$ 3.23\\
2014 Oct 24 & M4.0 & [80, -410]& 07:40:59 &11.6 $\pm$ 1.2& 1.05&1.7 $\pm$ 1.5&0.28$\pm$ 0.25& 4.16 $\pm$ 0.20 &4.14 $\pm$ 0.42\\
\label{table1}
\enddata
\end{deluxetable}

\end{document}